\documentclass[
  prb,
  superscriptaddress,
  reprint
  ]{revtex4-1}
\usepackage[utf8]{inputenc}
\usepackage{CJKutf8}
\usepackage[T1]{fontenc}
\usepackage[usenames, dvipsnames]{color}
\usepackage{graphicx}
\usepackage{float}
\usepackage{empheq}
\usepackage{textcomp}
\usepackage{bm}
\usepackage{verbatim}
\usepackage{amsmath}
\usepackage{mathrsfs} 
\usepackage{amssymb}
\usepackage{amsthm}
\usepackage{xfrac}
\usepackage{hyperref}
\usepackage{physics}
\usepackage{dsfont}
\graphicspath{ {images/} }

\newcommand*{\Neel}{}
\def\Neel/{N\'eel}
\newcommand*{\Schrodinger}{}
\def\Schrodinger/{Schr\"odinger}

\newcommand*{\zhat}{\ensuremath{\hat{z}}}

\newcommand*{\bmag}{\ensuremath{\mathbf{m}}}
\newcommand*{\bn}{\ensuremath{\mathbf{n}}}

\newcommand*{\bx}{\ensuremath{\mathbf{x}}}
\newcommand*{\br}{\ensuremath{\mathbf{r}}}
\newcommand*{\bA}{\ensuremath{\mathbf{A}}}
\newcommand*{\bB}{\ensuremath{\mathbf{B}}}

\newcommand*{\bX}{\ensuremath{\mathbf{X}}}
\newcommand*{\bR}{\ensuremath{\mathbf{R}}}

\newcommand*{\ma}{\ensuremath{\mathbf{m}_A}}
\newcommand*{\mb}{\ensuremath{\mathbf{m}_B}}

\newcommand*{\bEta}{\ensuremath{\bm{\eta}}}

\newcommand*{\bk}{\ensuremath{\mathbf{k}}}

\definecolor{basebg}{RGB}{0, 43, 54}
\definecolor{basefg}{RGB}{131, 148, 150}
\definecolor{colormmablue}{RGB}{94, 129, 181}
\definecolor{colormmaorange}{RGB}{225, 156, 36}
\definecolor{colormmagreen}{RGB}{143, 176, 50}
\begin{document}

\title{Topological spin Hall effects and tunable skyrmion Hall effects
  in uniaxial antiferromagnetic insulators}

\begin{CJK*}{UTF8}{gbsn}
\author{Matthew W.~Daniels} 
  \email[Corresponding author:~]{matthew.daniels@nist.gov}
  \affiliation{ Department of Physics,
                Carnegie Mellon University,
                Pittsburgh, PA 15213, USA }
  \affiliation{ Institute for Research in Electronics and Applied Physics,
                University of Maryland, College Park,
                MD 20742, USA }

\author{Weichao Yu (余伟超)}
\affiliation{ Department of Physics and State Key
              Laboratory of Surface Physics,
              Fudan University,
              Shanghai 200433, China}

\author{Ran Cheng}
  \affiliation{ Department of Physics,
                Carnegie Mellon University,
                Pittsburgh, PA 15213, USA }
  \affiliation{ Department of Electrical and Computer Engineering,
                University of California, Riverside, CA 92521, USA}

\author{Jiang Xiao (萧江)}
\affiliation{ Department of Physics and State Key
              Laboratory of Surface Physics,
              Fudan University,
              Shanghai 200433, China}
\affiliation{ Institute for Nanoelectronics Devices and Quantum Computing, Fudan University, Shanghai 200433, China }

\author{Di Xiao}
  \affiliation{ Department of Physics,
                Carnegie Mellon University,
                Pittsburgh, PA 15213, USA }

\begin{abstract}
  Recent advances in the physics of current-driven antiferromagnetic
  skyrmions have observed the absence of a Magnus force. We outline
  the symmetry reasons for this phenomenon, and show that this
  cancellation will fail in the case of spin polarized
  currents. Pairing micromagnetic simulations with semiclassical spin
  wave transport theory, we demonstrate that skyrmions produce a
  spin-polarized transverse magnon current, and that spin-polarized
  magnon currents can in turn produce transverse motion of
  antiferromagnetic skyrmions. We examine qualitative differences in
  the frequency dependence of the skyrmion Hall angle between
  ferromagnetic and antiferromagnetic cases, and close by proposing a
  simple skyrmion-based magnonic device for demultiplexing of spin
  channels.
\end{abstract}

\maketitle
\end{CJK*}

\section{Introduction}
\label{sec:intro}
Skyrmions have long been appreciated in the magnetism community for
their topological stability\cite{Belavin:1975uk} and soiltonic
dynamics.\cite{Thiele:1973aa} In metallic ferromagnets, skyrmions
cause a transverse deflection of itinerant electrons known as the
topological Hall effect.\cite{Stone:1996aa} When the current is strong
enough to drive ferromagnetic skyrmions, the skyrmion itself undergoes
motion transverse to the current; this reciprocal process is the
skyrmion Hall effect. These two effects are related through
conservation of momentum.\cite{Tchernyshyov:2015fr} Similar effects
exist in insulating magnets, where magnons, rather than electrons,
represent the low-lying excitations of the
system.\cite{Kong:2013aa,Iwasaki:2014aa,Schutte:2014aa}

Recently, skyrmions in metallic antiferromagnets have garnered
attention as an attractive alternative to their ferromagnetic
counterparts. Unlike ferromagnetic skyrmions, antiferromagnetic
skyrmions do not undergo transverse motion in response to electronic
current.\cite{Barker:2016fq,Zhang:2016gy,song:2016aa} The crucial
difference is the two-sublattice structure of the antiferromagnet:
though each ferromagnetic sublattice nominally experiences a Magnus
force, there is a perfect cancellation between the two, and the
skyrmion moves only longitudinally with the current. This absence of
the skyrmion Hall effect in antiferromagnets has been lauded for the
resulting simplicity of the skyrmion dynamics. It may represent a
significant technological advantage in the quest for spin texture
based applications such as racetrack memory\cite{zhang:2016aa} or
skyrmion computing.

To date, less has been said about antiferromagnetic skyrmions in
insulating systems, and whether or not skyrmion Hall effects arise
therein. In this paper, we explore how a magnon/skyrmion \emph{spin}
Hall effect appears in antiferromagnetic magnon-skyrmion
interactions. A crucial feature of spin waves in easy axis
antiferromagnets is that they are spin-polarized in general. This is
expected theoretically,\cite{Keffer:1952jw,Cheng:2014aa} and such
magnon-mediated spin currents have recently been observed
experimentally in materials such as $\text{Cr}_2\text{O}_3$ and
$\text{MnPS}_3$.\cite{Seki:2015es,shiomi:2017aa}
A consequence of these spin polarized currents, and the topic of the
present article, is that the perfect cancellation of Magnus forces
discussed in previous works
\cite{Barker:2016fq,Velkov:2016uk,Zhang:2016gy,Zhang:2016kf} does not
always hold in magnonics. We present theory and simulation
demonstrating how one can generate magnonic forces on
antiferromagnetic skyrmions, and how, conversely, one can use
skyrmions as spin-splitters in antiferromagnetic magnon devices.

In Sec.~\ref{sec:symmetry-considerations}, we justify these claims by
presenting general symmetry arguments. In Sec.~\ref{sec:lagrangian},
we discuss the theoretical framework underlying the rest of the paper.
Sec.~\ref{sec:magnon-hall} presents semiclassical transport results,
supported by simulation, to describe the magnonic topological spin
Hall effect generated by antiferromagnetic skyrmions.
Sec.~\ref{sec:skyrmion-cc} derives the equations of motion of the
skyrmion texture in this coupled system, demonstrating in simulation
an angle-tunable skyrmion Hall effect. We conclude in
Sec.~\ref{sec:applications} by describing some possible applications
of these effects in magnonic logic.

\section{Symmetry considerations and summary of results}
\label{sec:symmetry-considerations}
\begin{figure}
  \centering
  \includegraphics[width=\columnwidth]{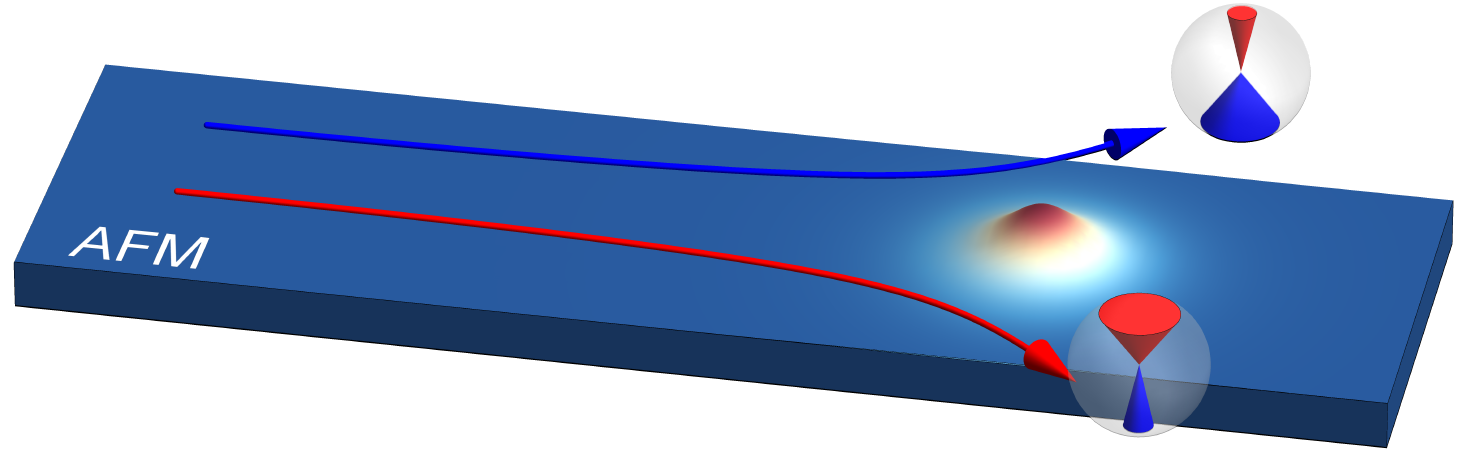}
  \caption{This diagram shows a skyrmion, far right, subject to spin
    unpolarized magnon current entering from the far left. The blue
    and red arrows representing left- and right-handed spin wave
    channels have equal amplitude. Mirror symmetry about the major
    axis of the strip is preserved, since the $m_z$ carried by each
    spin wave channel changes sign under the mirror operation. The
    presence of this mirror symmetry protects the skyrmion from any
    forces transverse to the mirror symmetry axis. If the spin
    accumulation becomes asymmetric across this axis (as in the left
    and right panels of Fig.~\ref{fig:magnon-hall-figs}), then a
    transverse skyrmion Hall will necessarily manifest.}
  \label{fig:symmetry}
\end{figure}
We wish to study in antiferromagnets how spin waves interact with
skyrmion configurations of the antiferromagnetic order parameter.  A
skyrmion in a 3-dimensional\footnote{The order parameter is embedded
  in 3D, but in principle it must have unit magnitude or nearly unit
  magnitude to give sensible results.} field $\bn(\br)$ over a
2-dimensional plane is characterized by two criteria: it has a finite
and nonzero characteristic length scale;\cite{Manton:vEuV_4gy} and it
has nonzero topological charge,
\begin{equation}
  \label{eq:skyrmion-num-def}
  Q = \frac{1}{4\pi}\int_{\mathbb{R}^2} \hat n \cdot \left( \frac{\partial \hat n}{\partial x} \cross \frac{\partial\hat n}{\partial y} \right)\dd{x}\dd{y}.
\end{equation}
Under the assumption that $\hat n$ has a uniform boundary condition at
infinity, it is straightforward to show that $Q$ is quantized and that
it labels topologically protected configurations of
$\hat n$.\cite{Belavin:1975uk} In ferromagnetic spintronics, the unit
vector $\hat n$ in Eq.~\eqref{eq:skyrmion-num-def} is taken to be the
local magnetization vector.\cite{Nagaosa:2013aa} In antiferromagnets,
we instead take $\hat n = \bn/\Vert\bn\Vert$ to be the normalized
staggered order vector $\bn = (\ma - \mb)/2$, defined as the local
difference between sublattice magnetizations.\cite{Velkov:2016gy}
Though skyrmion-like bubble states can be stabilized by dipolar
interactions in ferromagnets, we obviously lack this mechanism in
antiferromagnets. Here, and in many ferromagnets of interest as well,
our skyrmions are stabilized entirely by the competition between
magnetocrystalline anisotropy and the Dzyaloshinskii-Moriya
interaction in the presence of the exchange interaction. The
Dzyaloshinskii-Moriya interaction arises due to inversion symmetry
breaking,\cite{Moriya:1960go} either in the crystal lattice itself or
due to interfaces between the magnet and another material. Because the
antiferromagnetic free energy in terms of $\bn$ has a similar form at
leading order to the ferromagnetic free energy in $\bmag$, many
results on the stationary stabilization of ferromagnetic
skyrmions\cite{Rohart:2016fw,Rohart:2013ef} are expected to translate
qualitatively to the antiferromagnetic case.

In ferromagnets, it is known that the emergent magnetic field of a
skyrmion produces a transverse force on electronic current. In recent
literature on antiferromagnetic skyrmions, the absence of an observed
skyrmion Hall effect has often been explained by a simple model
treating a ferromagnetic skyrmion per
sublattice.\cite{Barker:2016fq,Zhang:2016gy,Zhang:2016kf} In this
picture, the two ferromagnetic skyrmions have opposite skyrmion
numbers and opposite emergent magnetic fields. Each ferromagnetic
skyrmion thereby feels an oppositely signed Magnus force, and as the
two are bound together by exchange coupling, the full
antiferromagnetic skyrmion feels no net transverse force at
all.

Despite this cancellation of transverse forces acting on the skyrmion,
there is no reason to believe that individual electrons feel no
transverse force. A force perpendicular to a particle's trajectory
represents the breaking of mirror symmetry across the plane normal to
that force. A particle that couples to a spin texture will see this
mirror symmetry broken in the presence of a skyrmion, and so we expect
a transverse force to appear in general. Semiclassical transport
theory predicts\cite{Cheng:2012kl} just such an emergent Lorentz force
on electrons due to the emergent magnetic field
$\hat n \cdot (\partial_x \hat n \times \partial_y \hat n)$ of an
antiferromagnetic skyrmion, and recent first principles calculations
have suggested the existence of a spin Hall effect for electrons
flowing through metallic antiferromagnetic
skyrmions.\cite{Buhl:2017wf} Similar suggestions are also starting to
be understood in ferrimagnetic systems near
compensation.\cite{PhysRevLett.122.057204}

Yet so long as an equal number of spin up and spin down carriers
participate in such a spin Hall effect, mirror symmetry of the spin
accumulation across the direction of current flow is preserved. This
also corresponds to a mirror symmetry of the spin current channels, as
in Fig.~\ref{fig:symmetry}. As such, the skyrmion itself is protected
by symmetry from experiencing transverse forces; any such force would
break the mirror symmetry. This explains the absence of a skyrmion
Hall effect as observed in antiferromagnetic
simulations,\cite{Zhang:2016kf} but we emphasize that the symmetry
protection holds only when the incoming current is spin
unpolarized.\footnote{By \emph{spin unpolarized}, we mean simply that
  the spin current operator can be factored into its spin and velocity
  components. A current where both spin channels have the same number
  current distribution is spin unpolarized by our definition.} We also
note that a proper analysis of the spin current or spin accumulation
distribution has both a vector nature (in the current or position
component) and pseudovector or axial nature (in the spin
component).\cite{shi:2006} Symmetry analysis that attempts to use the
``mean'' spin current in one direction or another will give incorrect
results in the general case.

Magnonic excitations are also subject to these symmetry arguments;
since a skyrmion breaks the mirror symmetry of the spin texture,
magnons passing through one undergo a topological spin Hall effect. We
have predicted the form of this force in semiclassical transport in
Ref.~\onlinecite{daniels:2018nonabelian}; the result is qualitatively
similar to the electron case, and involves coupling to the emergent
magnetic field of the skyrmion.

Like electrons' spin degree of freedom, antiferromagnetic spin waves
carry a spinor-valued internal degree of freedom called the magnon
isospin.\cite{daniels:2018nonabelian} It arises due to the two-fold
sublattice structure, and characterizes the polarization and
handedness of the staggered order precession.\cite{Keffer:1952jw} Its
$z$-component on the Bloch sphere is referred to alternatively as the
spin wave chirality or the isospin charge, and has shown to be
explicitly connected with the breaking of mirror
symmetry.\cite{proskurin:2017aa} This chirality can be associated with
magnonic spin currents in easy axis
antiferromagnets,\cite{Cheng:2014aa,daniels:2018nonabelian} and
tunably generated through spin-transfer torques,\cite{Daniels:2015aa}
circularly polarized light,\cite{proskurin:2018:excitation} or
oscillating applied magnetic fields.\cite{Cheng:2016kv}

For antiferromagnetic skyrmions driven by isospin-charged spin
waves---that is, those with a circular or elliptical polarization---we
expect the mirror symmetry restricting transverse forces to be
broken. The consequence, which we show explicitly in the remainder of
the paper, is that the skyrmion experiences transverse forces
proportional to the degree of mirror symmetry breaking. And though the
anisotropy that associates isospin charge to spin current is necessary
to stabilize the skyrmion ground state, we note that the net spin
carried by spin waves is immaterial to the skyrmion Hall effect from a
symmetry perspective---the handedness of spin wave precession alone is
enough to break mirror symmetry across the skyrmion trajectory and
induce transverse forces.

Finally, though we are not aware of any prior demonstrations of the
effect in the literature, this symmetry argument also suggests that a
spin-polarized electron current in an antiferromagnetic metal should
produce a skyrmion Hall effect similar to the magnon-driven one
we explore in the remainder of this paper. Detailed analysis of this
prediction for electronic systems is left to future research.

\section{Formulation}
\label{sec:lagrangian}

In this paper, we work with 2D collinear antiferromagnets with
uniaxial anisotropy and broken inversion symmetry. The latter leads to
a Dzyaloshinskii-Moriya interaction. In the case of a bipartite,
$g$-type antiferromagnet, these terms lead collectively to a free
energy
\begin{equation}
  \label{eq:hamiltonian}
  F = \int \left[\mathcal F_J + \mathcal F_D + \mathcal F_K\right]\,\text{d}^2x.
\end{equation}
Each free energy density is given by
\begin{subequations}
  \begin{align}
    \label{eq:ham-densities}
    \mathcal F_J &= \frac{Z}{2}\ma\cdot\mb - \frac{J}{2}\nabla\ma \cdot \nabla \mb,\\
    \mathcal F_D &= \frac{D}{2}\left[\ma \cdot (\nabla\times\mb)+\mb\cdot(\nabla\times\ma)\right],\\
    \mathcal F_K &= -\frac{K}{2}\left[ (\ma\cdot\zhat)^2 + (\mb\cdot\zhat)^2\right],
  \end{align}
\end{subequations}
where $\ma(\br,t)$ and $\mb(\br,t)$ are the continuum representation
of the $A$- and $B$-sublattice magnetic orders. These are normalized
vectors, with the saturation magnetization $M_s$ on each sublattice
absorbed into the interaction coefficients. The antiferromagnetic
exchange energy $J$ is that of the underlying lattice Hamiltonian
$\sum_{\langle i j\rangle}J\bmag_i\cdot\bmag_j$, while $Z$ is the
exchange energy density. $K$ gives the strength of the uniaxial
magnetocrystalline anisotropy. We have expressed the
Dzyaloshinskii-Moriya interaction with strength $D$ in the form that
arises from bulk inversion symmetry breaking; substituting
$\nabla \mapsto \hat z \times \nabla$ in this term recovers the
interfacial version of the interaction with no qualitative change in
the dynamics.\cite{daniels:2018nonabelian}

We will assume that the ground state of the system is collinear and
antiferromagnetic, with the ground state $\ma^0 = \zhat$ and
$\mb^0=-\zhat$ for concreteness. This constrains the value of $D$ to
lie below the critical value that would lead to a spiral
state.\cite{Rohart:2013ef} We present this model for consistency with
other theoretical literature, but our simulation results and many
important applications for antiferromagnetic skyrmions take place in
synthetic antiferromagnets. Our model also captures these systems with
only small quantitative adjustments.\footnote{The crucial adjustments
  in moving from a bipartite to a synthetic antiferromagnet are to
  send
  $-|J_\text{AFM}|\nabla\ma \cdot\nabla\mb \mapsto
  (|J_\text{FM}|/2)(|\nabla\ma |^2+|\nabla\mb |^2)$ and
  $(D/2)(\ma \cdot\nabla\times\mb +\mb \cdot\nabla\times\ma ) \mapsto
  (D/2)(\ma \cdot\nabla\times\ma + \mb \cdot\nabla\times\mb )$. In the
  staggered order language, the former amounts to changing the sign of
  $\nabla\bmag\cdot\nabla\bmag$ in the exchange free energy, and the
  latter amounts to changing the sign of $\bn\cdot(\nabla\times\bn)$
  in the DMI free energy. Except for details of the isospin
  Hamiltonian $\mathscr{H}$ in Eq.~\eqref{eq:isospin-eom} arising from
  the presence of time-reversal plus sublattice exchange symmetry in
  synthetic antiferromagnets (which is lost in textured bipartite
  systems), the semiclassical magnon equations of motion
  \eqref{eq:semiclassical-eqs} are structurally identical between the
  two cases.\cite{daniels:2018nonabelian}}

To treat magnon-skyrmion interactions, we implement the formalism laid
out in Ref.~\onlinecite{daniels:2018nonabelian}. The magnetization
fields $\ma$ and $\mb$ are decomposed into their fast and slow modes,
which represent magnon and skyrmion dynamics,
respectively.\cite{Kong:2013aa} This separation is valid so long as
the spin texture can be modeled as passing through quasistatic
equilibria of the magnetic free energy, though at high speeds emergent
relativistic effects can become important,\cite{Kim:2014aa} which we
do not model here. The magnon modes are collected into a 4D vector and
obey a Schr\"odinger-like equation of motion. The spin texture
information of the staggered order $\bn(\br)=(\theta(\br),\phi(\br))$
is represented by an emergent magnetic field
$\bm B= B\hat z = \hat z \,\sin\theta(\nabla\theta\times\nabla\phi)$
corresponding to the integrand of
Eq.~\eqref{eq:skyrmion-num-def}.\cite{Nagaosa:2013aa} Unlike in
Ref.~\onlinecite{daniels:2018nonabelian}, we preserve Lagrangian terms
at zeroth order in the spin wave fields, which will give the inertial
response of the spin texture. The details of this sector of our
Lagrangian are equivalent to the Lagrangian underlying
Ref.~\onlinecite{Tveten:2013aa}.
\begin{figure*}
  \centering
  \includegraphics[width=0.32\textwidth]{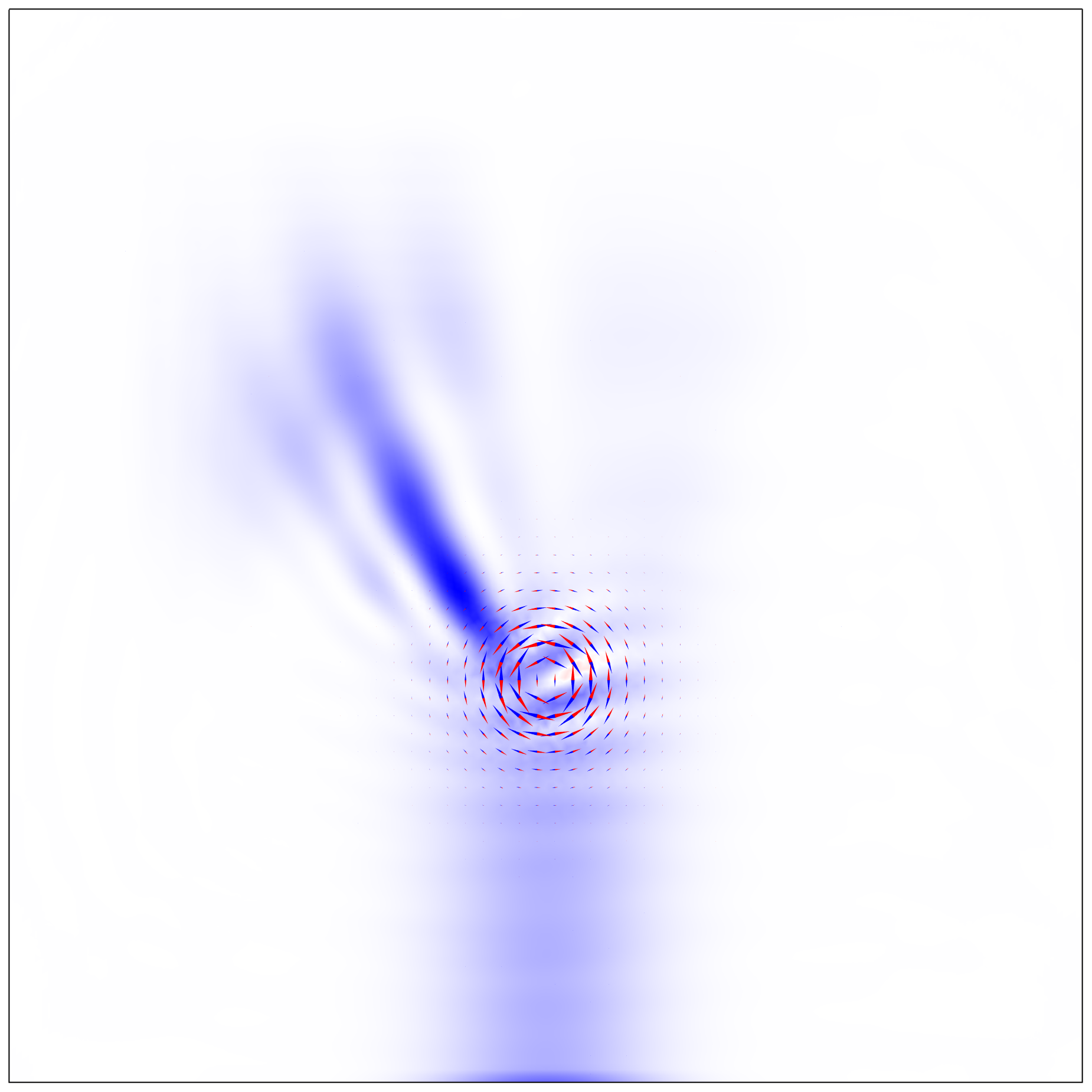}
  \includegraphics[width=0.32\textwidth]{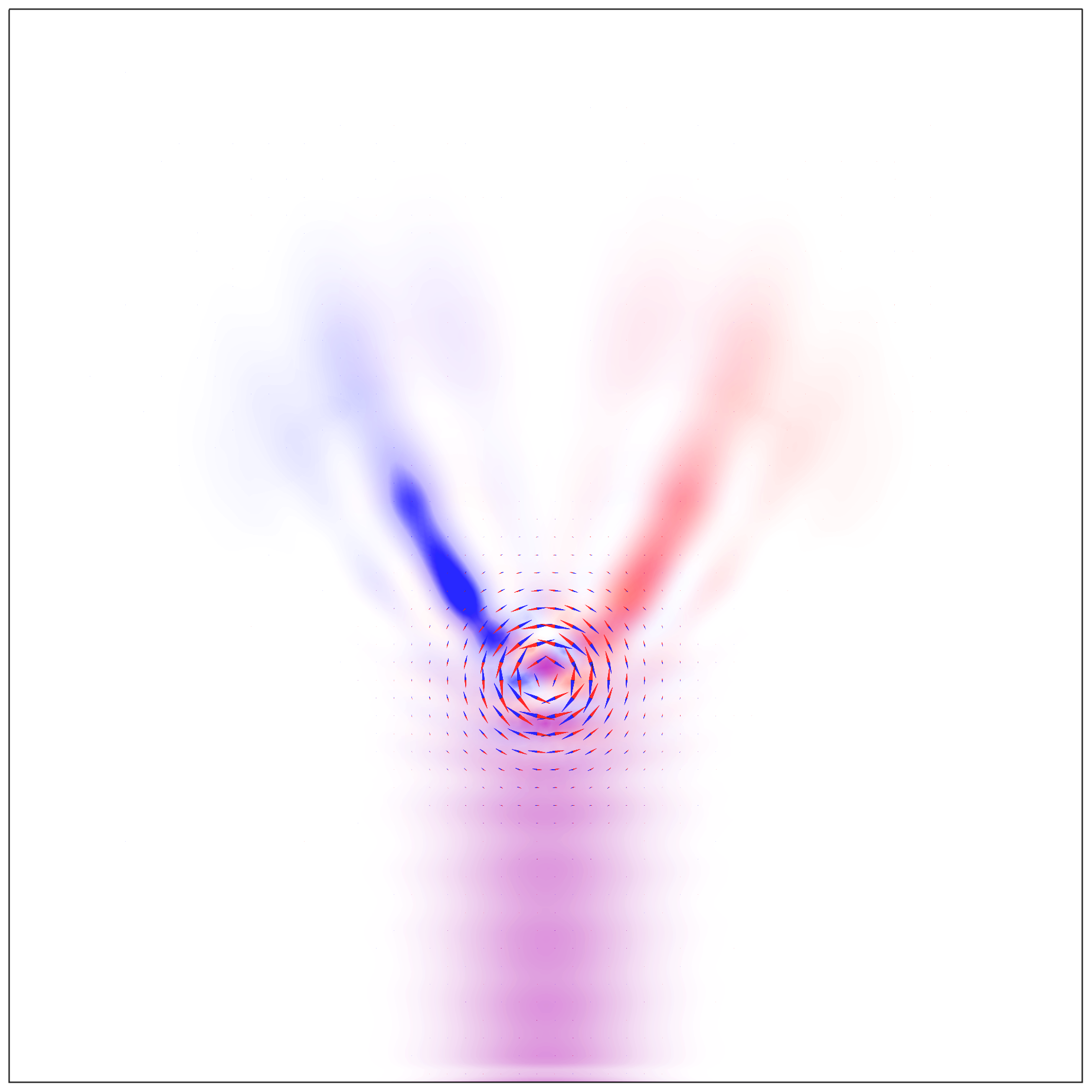}
  \includegraphics[width=0.32\textwidth]{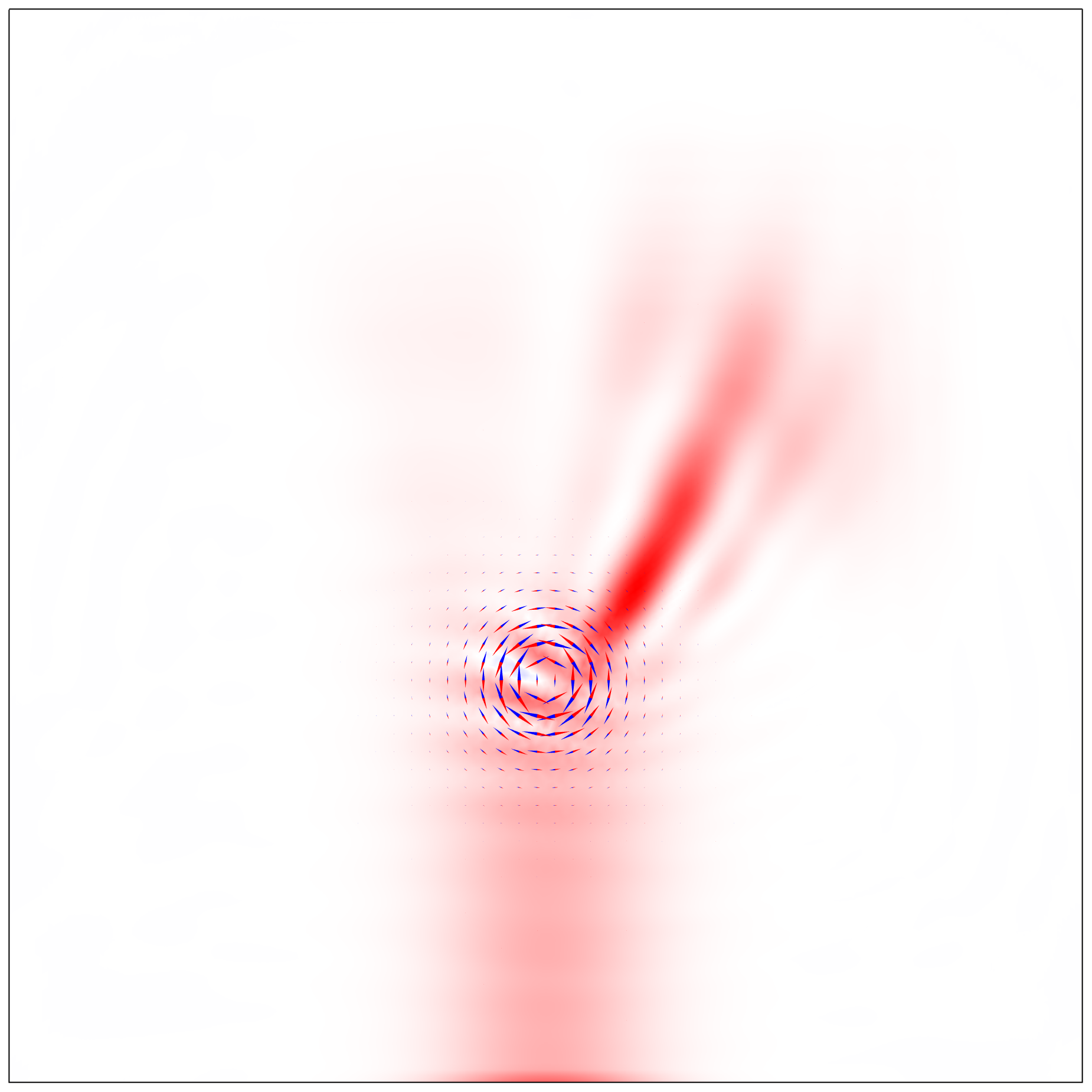}
  \caption{Micromagnetic simulations of the magnonic topological spin
    Hall effect in antiferromagnetic skyrmions. Blue and red represent
    left and right handed spin wave precession of the local N\'eel
    order about equilbrium, plotted as the square of the spin wave
    amplitude integrated over the entire simulation time. In all
    cases, a fluctuating magnetic field injects coherent spin waves
    ($\omega/2\pi=30$ GHz) at the bottom of the frame. Left:
    left-handed spin waves are deflected to the left. Right:
    right-handed spin waves are deflected to the right. Center:
    linearly polarized spin waves, an equal superposition of the left-
    and right-handed magnon bands, are decomposed by the skyrmion into
    its spin polarized eigenmodes. The sample size is
    $800\,\text{nm}\times 800\,\text{nm}$.}
  \label{fig:magnon-hall-figs}
\end{figure*}

\section{Skyrmion-induced magnon Hall effect}
\label{sec:magnon-hall}



In Ref.~\onlinecite{daniels:2018nonabelian}, we have recently worked
out the general semiclassical transport theory for antiferromagnetic
spin waves in a system such as that defined by
Eq.~\eqref{eq:hamiltonian}. Here we specialize that result to the
skyrmion case. Recall that there are two degenerate spin wave modes in
the uniform easy-axis antiferromagnet, which correspond to right- and
left-handed precessions of the staggered order. A coherent magnonic
excitation can be indexed by a spinor quantity $\bEta$ quantifying the
linear combination of right- [$\bEta = (1,0)$] and left-handed
[$\bEta=(0,1)$] components. We take $\bEta$, the magnon isospin, to be
normalized for a single magnon wave-packet. When the right- and
left-handed magnon frequency bands are degenerate, $\bEta$ indexes
that degenerate subspace and can explore the Bloch sphere with no
energy cost. The addition of the Dzyaloshinskii-Moriya interaction
breaks this degeneracy, but only weakly, so certain rotations of
$\bEta$ incur small energy costs.\cite{culcer:2005coherent,shindou:2005aa}

We will assume that the skyrmion in our problem is rigid---that is,
its shape is fixed but its position may vary. This assumption implies
that time dependence of the spin texture can be factored through the
rigid skyrmion's position coordinate $\bR=(X(t),Y(t))$. The
semiclassical Lagrangian for a single magnon wave-packet is given by
\begin{equation}
  L_\text{WP} = -\bEta^\dagger\left[ \sigma_z (\dot{\bA}\cdot\br-V)-\dot{\bk}\cdot\br-\mathscr H\right]\bEta,
\end{equation}
with $\br$ and $\bk$ the wave-packet's phase space variables,
$\sigma_z$ the usual Pauli matrix, and $\bm{A}(\br,t)$ and $V(\br,t)$
vector and scalar potential functions underlying the emergent
electromagnetic
fields.\cite{Dugaev:2005aa,Guslienko:2010kq,Nagaosa:2013aa} The vector
potential is related to the skyrmion's emergent magnetic field through
$\bm{B} = \nabla\times\bm A$. The $2\times 2$ isospin Hamiltonian
$\mathscr{H}$ is derived by projecting out the negative frequency
subspace of the $4\times 4$ spin wave Hamiltonian governing the
$\hat \theta$ and $\hat \phi$ components of $\ma$ and
$\mb$.\cite{daniels:2018nonabelian} With the time dependence of the
electromagnetic potentials replaced with the collective coordinate
dynamics of $\bm{R}$, the equations of motion for the magnon
wavepackets become
\begin{subequations}
  \label{eq:semiclassical-eqs}
\begin{align}
  \dot{\bk} &= \chi (\dot{\br} - \dot{\bR})\cross \bB - \frac{\partial \omega}{\partial \br},\label{eq:bq-dot}\\
  \dot{\br} &= \frac{\partial \omega}{\partial \bk},\label{eq:bx-dot}\\
  \text{and}\quad i\dot{\bEta} &= \mathscr H \bEta,\label{eq:isospin-eom}
\end{align}
\end{subequations}
at second order in derivatives of the spin texture.\footnote{In
  certain cases where the DMI vector is out of plane, Berry curvature
  terms may appear in the momentum space equation above. There are
  explored elsewhere in the
  literature;\cite{Cheng:2016ku,zyuzin:2016aa} so here we neglect them
  and we focus only on the skyrmion-specific phenomenology.}  Here
$\omega=\omega(\br)$ is the spin wave frequency of the local
magnon band structure and $\chi=\bEta^\dagger\sigma_z\bEta$ is the
emergent charge.

The isospin charge $\chi$ corresponds to the handedness of the spin
wave: it is $\pm 1$ for right- and left-handed waves, and vanishes for
linearly polarized waves. That it acts as the coupling constant
between magnon and skyrmion reflects our symmetry arguments from
Sec.~\ref{sec:symmetry-considerations}: the degree and sign of mirror
symmetry breaking controls the transverse force on the magnon. We
expect from Eq.~\eqref{eq:bq-dot} that right- and left-handed waves
experience oppositely signed Lorentz forces in the presence of a
skyrmion spin texture.

To verify this prediction, we carried out micromagnetic simulations of
a skyrmion in a synthetic antiferromagnet. We numerically solved the
Landau-Lifshitz-Gilbert equation using finite elements methods
code.\cite{Lan:2017aa,lan:2015:diode,Yu:2016ut,yu:2018:polarization}
The parameters are chosen following
Refs.~\onlinecite{wang:2011:all-magnonic,lan:2015:diode}: gyromagnetic
ratio $\gamma = 2.21 \times 10^5\; \text{Hz}/(\text{A}/\text{m})$,
exchange stiffness
$\mathcal J = 4.00 \times 10^{-12}\;\text{J}/\text{m}$,
Dzyaloshinskii-Moriya energy
$\mathcal D = 1.23 \times 10^{-4}\; \text{J}/\text{m}^2$, anisotropy energy
$\mathcal K = 4.73 \times 10^3 \;\text{J}/\text{m}^3$, and interlayer
antiferromagnetic energy
$\mathcal Z = 6.09 \times 10^4\; \text{J}/\text{m}^3$. We set the
Gilbert damping to $\alpha = 10^{-4}.$ In these simulations, we
generated either circularly polarized spin waves, by applying a
rotating magnetic field
$\bm{h}_\text{RH/LH} = h_0 \left[ \sin(\omega t) \hat x \mp
  \cos(\omega t) \hat y \right]$, or linearly polarized spin waves, by
applying $\bm{h}_\text{X}=h_0 \sin\omega t \hat y$ or
$\bm{h}_\text{Y}=h_0 \sin\omega t \hat x$ for X- or Y-polarizations.

The results are displayed in Fig.~\ref{fig:magnon-hall-figs}. We find
that scattering right-handed ($\chi=+1$) and left-handed ($\chi=-1$)
spin waves from the skyrmion results in right-ward and left-ward
deflection, respectively. Leftward deflection of right-handed magnons
from a skyrmion is consistent with the ferromagnetic skyrmion Hall
effect;\cite{Iwasaki:2014aa,Schutte:2014aa} as left-handed magnons do
not exist in ferromagnets,\footnote{Left-handed spin waves can be
  found in ferromagnets in principle, but not as low-lying excitations
  at the energy scales of interest here.} we find that this latter
effect is a uniquely antiferromagnetic phenomenon. When linearly
polarized spin waves ($\chi=0$) are injected upon the skyrmion
texture, the skyrmion splits the signal into its two spin polarized
channels, producing a transverse spin current with no net transverse
magnon number current. This result for linearly polarized waves is
analogous from a symmetry perspective to the electronic topological
spin Hall effect predicted in Ref.~\onlinecite{Buhl:2017wf}.

While the spin waves are strongly defected by the Lorentz force in
Eq.~\eqref{eq:bq-dot}, their isospin degree of freedom does not remain
static throughout this process. According to
Eq.~\eqref{eq:isospin-eom}, the isospin vector itself undergoes
dynamics in the presence of a nontrivial $\mathscr H$, such as arises
within the skyrmion texture. This in turn affects the isospin charge
$\chi$. Even supposing that a purely right-handed spin wave signal
enters the skyrmion, it will constantly undergo some isospin dynamics,
resulting in the production of some signal in the left-handed
channel. Consequently, the nearly right-hand polarized signal exiting
the skyrmion picks up some ellipticity in
Fig.~\ref{fig:magnon-hall-figs}. Though our simulations show signs of
this behavior, the effect is small; it is also difficult to isolate,
occurring mainly near the skyrmion core. Attempting to incorporate
these dynamics into the equation for the Lorentz force indicates an
angular momentum transfer between magnon isospin and mechanical
angular momentum of the skyrmion, but only at higher orders in
perturbation theory than Eqs.~\eqref{eq:semiclassical-eqs} can
express. We expect that any effect on the dynamics of magnon and
skyrmion trajectories is negligible.

\begin{figure}
  \centering
  \includegraphics[width=0.49\columnwidth]{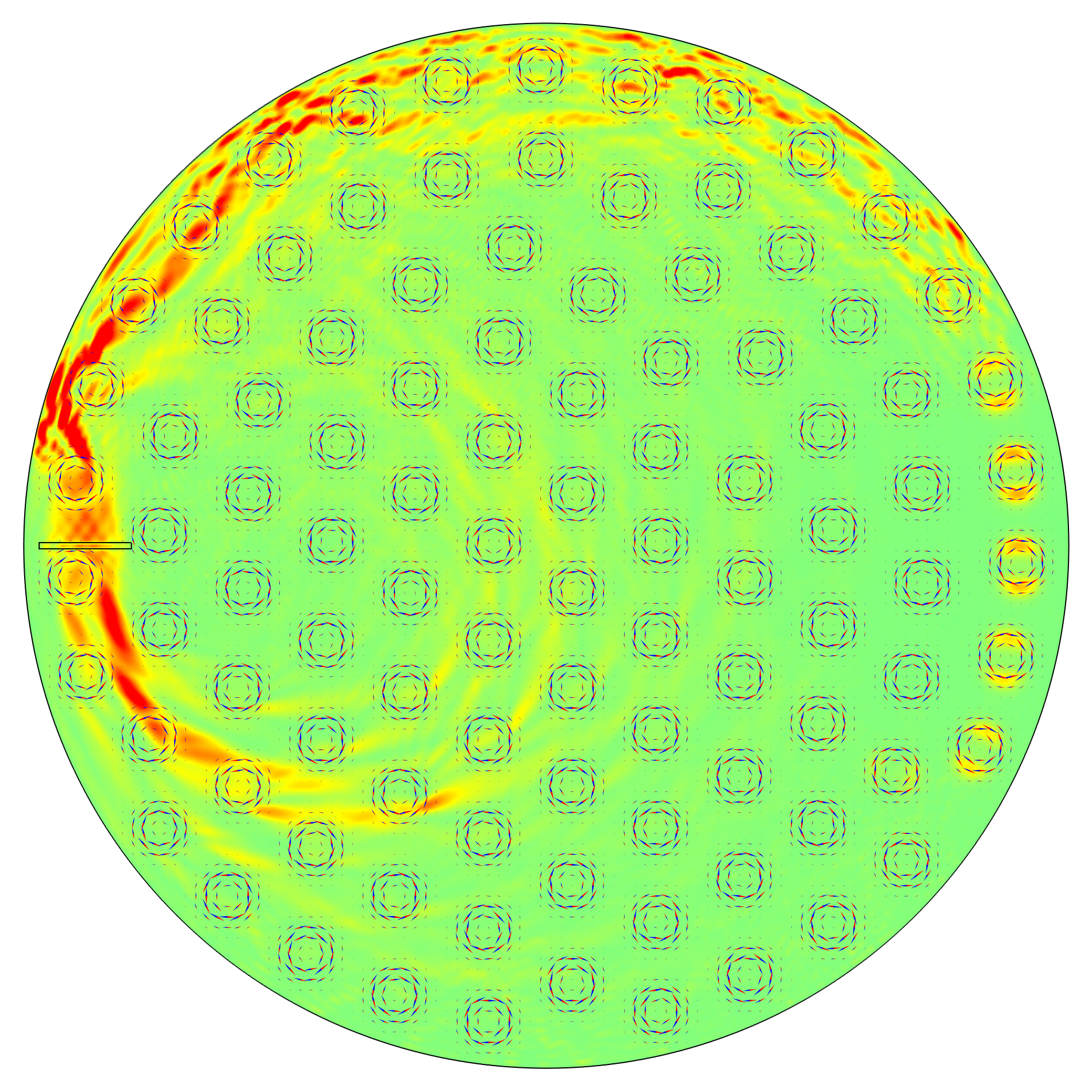}
  \includegraphics[width=0.49\columnwidth]{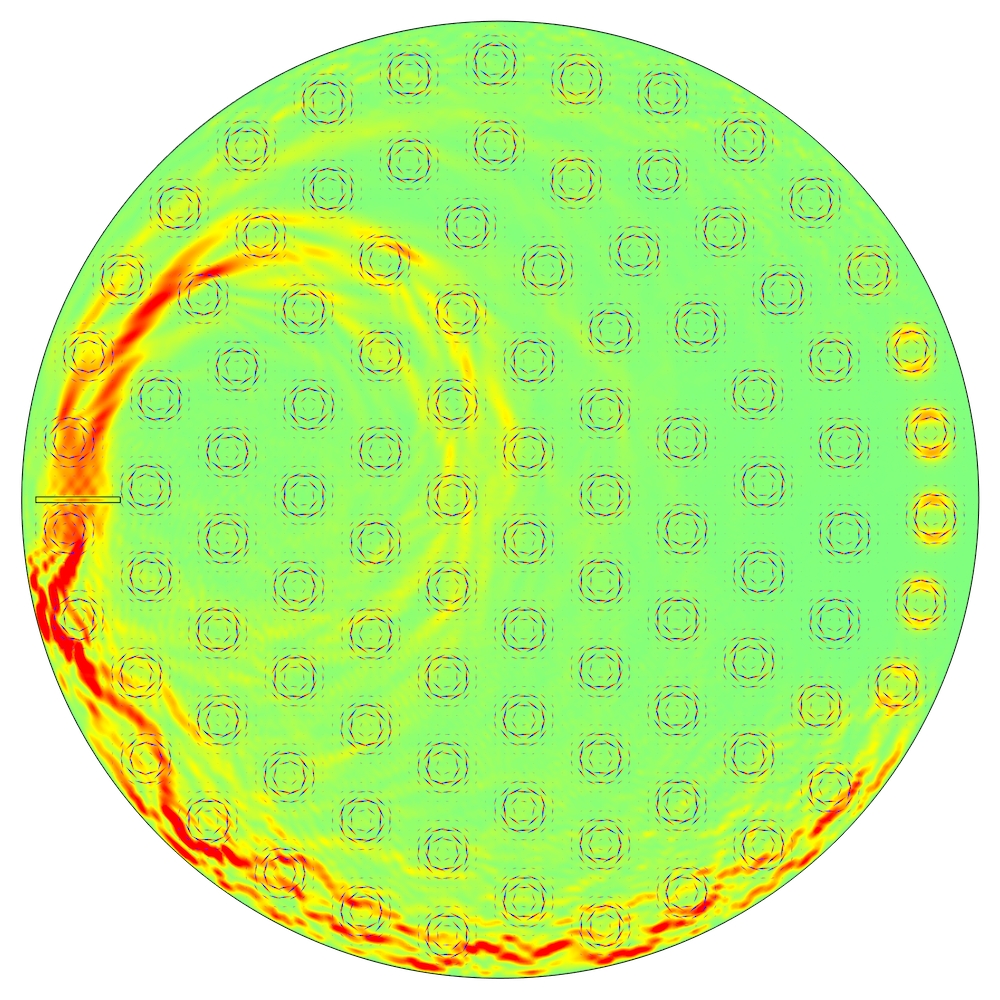}
  \caption{Edge modes of the antiferromagnetic skyrmion crystal. The
    color scale gives the squared spin wave amplitude integrated over
    the full simulation time. A simulated microwave antenna sits above
    the nanodisk on left side of the sample, indicated by a horizontal
    line.  Left and right-handed spin waves are injected at 20 GHz in
    figures left and right, respectively. As in the common toy model
    of the quantum spin Hall effect, cyclotron orbits of the magnons
    combined with an effective confining potential at the edge
    restrict the magnon trajectories to chiral directions
    corresponding to their isospin. The sample diameter is 1300 nm,
    and the snapshot was taken 2.1 ns into the simulation. Gilbert
    damping causes the spin wave intensity to decay away from the
    antenna.\label{fig:edge-modes}}
\end{figure}

Since antiferromagnetic skyrmions present a real-space Berry
curvature, magnons act as charged particles in the presence of a
perpendicular magnetic field. In a finite geometry, one might
therefore expect chiral edge modes in an antiferromagnetic skyrmion
crystal.  Using mircomagnetic simulations, we injected right- and
left-handed spin waves into a nanodisk with an artificially
constructed antiferromagnetic skyrmion lattice. The results are shown
in Fig.~\ref{fig:edge-modes}. Modes propagate away from the antenna in
the direction according to their handedness, decaying naturally via
Gilbert damping. A more quantiative analysis of this result has
recently been developed from lattice-level transport theory.\cite{PhysRevLett.122.187203}

\section{Skyrmion collective coordinate theory}

\label{sec:skyrmion-cc}
The spin wave dynamics and magnon-skyrmion interactions summarized in
the previous section are captured at the semiclassical level by a
wave-packet Lagrangian $L_{\text{WP}}$ that leads to
Eqs.~\eqref{eq:semiclassical-eqs}.\cite{daniels:2018nonabelian} A
Lagrangian for the full system can be obtained by adding to
$L_\text{WP}$ a sector detailing the inertial properties of the
skyrmion.

Write the density of magnons at $(\br,\bk)$ in phase space, with isospin
$\bEta$, as $\rho_{\br,\bk,\bEta}$. Accepting an uncertainty
$\Delta\br\Delta\bk \geq 2\pi$, we can expand the global wavefunction of magnons
into a linear combination of wavepackets. The total Lagrangian for all the
wavepackets together is merely their weighted sum
\begin{equation}
  \label{eq:psi-wp-decomp}
  L_\text{SWs}[\rho;\bB] = \int d^2\br\,d^2\bk\,d^2\bEta\;\rho_{\br,\bk,\bEta}L_{\text{WP}}[\br,\bk,\bEta,\bB].
\end{equation}
The uncertainty principle constraining this expansion means that the
following predictions will hold best for skyrmions large and smooth
compared to magnon wavelength. From Eq.~\eqref{eq:psi-wp-decomp} we
can immediately take a functional derivative with respect to $\bR(t)$
to obtain the magnonic force density over phase space. The result is
\begin{equation}
  \label{eq:magnon-force-1}
  \frac{\delta L_{\text{WP}}}{\delta \bR} = -\rho_{\br,\bk,\chi}\chi(\dot\br-\dot\bR)\times \bB_0(\br-\bR)
\end{equation}
where $\bB_0(\br-\bR) = \bB(\br)$. The total magnonic force on the skyrmion is
therefore
\begin{equation}
  \label{eq:magnon-force-total}
  \bm{F} = - \int d^2\br\,d^2\bk\,d^2\bEta\;\chi\rho_{\br,\bk,\chi}\left[  (\dot\br-\dot\bX)\times\bB_0(\br-\bR)\right],
\end{equation}
an isospin-charge-weighted sum of the reciprocal Lorentz forces by
each magnon. Since $\bB_0$ is radially symmetric, the component of
$\bm{F}$ perpendicular to the current flow
$\bm{j}\propto\langle \dot\br \rangle$ will vanish only if the isospin
charge distribution $\chi\rho_{\bx,\bk,\bEta}$ happens to be
asymmetric across $\bm{j}$. A special case of this condition occurs
when the incoming current is entirely unpolarized, as in the center
panel of Fig.~\ref{fig:magnon-hall-figs}, analogous to the electronic
case discussed widely in the literature and noted in
Sec.~\ref{sec:symmetry-considerations}.

\begin{figure}
  \centering
  \includegraphics[width=\columnwidth]{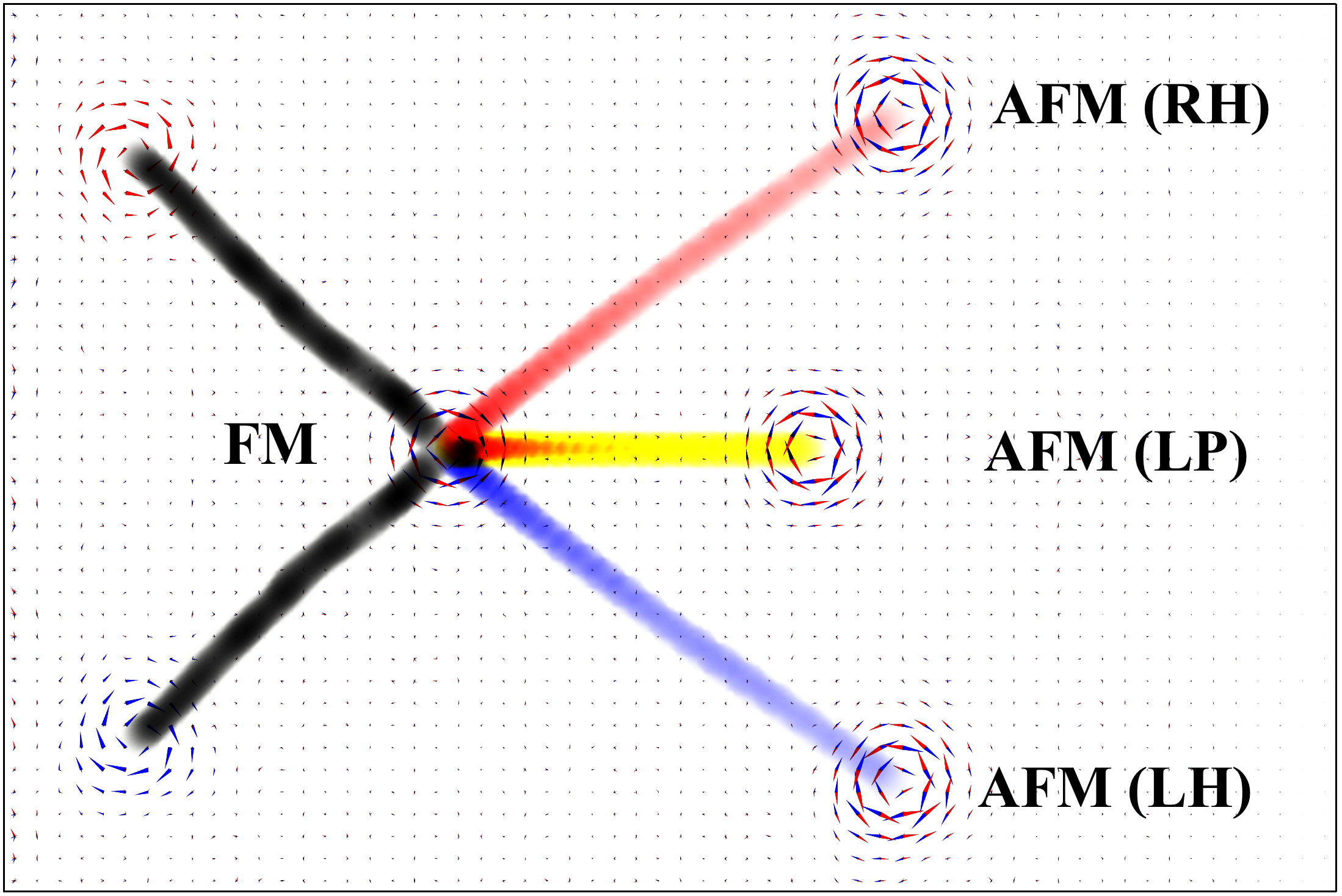}
  \caption{Overlaid skyrmion trajectories of multiple simulations in a
    $1200\,\text{nm}\times 800\,\text{nm}$ system.  In all cases, a
    spin wave current with wavenumber $k = 0.063\,\text{nm}^{-1}$
    flows from left to right. The three trajectories on the right
    correspond to nonzero interlayer coupling $J_\text{AF}\neq 0$,
    that is, a synthetic antiferromagnet. These trajectories are
    plotted at a snapshot time of $2.3\,\text{ns}$, and here
    $\omega(k) = 15.7\,\text{GHz}$. Yellow: an antiferromagnetic
    skyrmion is driven longitudinally, with no net Magnus force, by
    linearly polarized spin waves. Blue: an antiferromagnet skyrmion
    is driven to the right by left-handed spin waves. Red: an
    antiferromagnetic skyrmion is driven to the left by right-handed
    spin waves. The two black trajectores on the left correspond to
    $J_\text{AF}=0$, giving two decoupled ferromagets. Here, the
    sublattice skyrmions move transversely and counter-longitudinally
    to the magnon current, and the simulation required
    $100\,\text{ns}$ to reach this snapshot. In the ferromagnetic
    system, $\omega(k)=6\,\text{GHz}$.}
  \label{fig:skyrmion-trajs}
\end{figure}

Outside of this special case, the transverse force component has
considerable freedom. In Fig.~\ref{fig:skyrmion-trajs} we show
micromagnetic simulation results of antiferromagnetic skyrmion
trajectories driven by right, left, and linearly polarized spin
waves. There are two crucial differences from the ferromagnetic
case. First, the angle at which the skyrmion propagates can be tuned
by tuning the chirality of the driving magnon current. Second, because
the Lorentz force is the only force term present, and because
antiferromagnetic skyrmions have primarily diagonal mass tensors,
antiferromagnetic skyrmions move along, rather than against, the
magnon current driving them. The opposite is known to happen in
ferromagnetic skyrmion systems,\cite{Kong:2013aa} which we replicate
in Fig.~\ref{fig:skyrmion-trajs} by turning off the interlayer
coupling of our synthetic antiferromagnetic and allowing the two
sublattice skyrmions to behave as ferromagnetic skyrmions would. Their
reversed solutions are indicated in the figure by black trajectories.

\subsection{Magnon-mediated reduction in effective mass}
\label{sec:magn-medi-reduct}
The qualitative inertial behavior of antiferromagnetic skyrmions is
well-studied in the literature,\cite{Tveten:2013aa} so we do not dwell
on a detailed derivation here. As a rule and in the rigid skyrmion
approximation, antiferromagnetic skyrmions behave as classical
Newtonian point particles, complete with a mass term with its origins
in the sublattice interaction and the small magnetic moment carried by
the skyrmion. The non-interacting part of the skyrmion Lagrangian is
just
\begin{equation}
  \label{eq:lagrangian-cc-simpl}
  \mathcal L[\bR](\br,\bk) = \frac{1}{2}\tilde m_{ij}\dot R_i\dot R_j - \frac{S}{2}\omega\rho_n\epsilon^2 
\end{equation}
with the magnon number density
$\rho_n = \int d^2\bEta \,\rho_{\br,\bk,\bEta}$, the emergent skyrmion
mass density tensor
$\tilde m_{ij}(\br,\bk) = (S/\Omega)[ 2^{-1/2} -
\rho_n(\br,\bk)\omega(\br,\bk)/\Omega ]g_{ij}(\br)$, with the
characteristic frequency of the exchange interaction
$\Omega = Z\epsilon^2/S$, and $\epsilon$ the lattice constant. We
presume a square lattice for concreteness. The induced metric tensor
$g_{ij}$ on spin space is defined as
$g_{ij}= \partial_i\theta\partial_j\theta + \sin^2\theta
\partial_i\phi\partial_j\phi$.

The second term on the right-hand side of
Eq.~\eqref{eq:lagrangian-cc-simpl} is a potential energy landscape
presented to the skyrmion by the global magnon wavefunction. Though we
have claimed that this is the ``non-interacting'' part of the
Lagrangian, the presence of spin waves in the system softens the order
parameter of the skyrmion. This suggests that a skyrmion left to its
own devices would flow toward regions of high magnon density; this
matches with recent results indicating that antiferromagnetic
skyrmions will flow along the direction of a thermal
gradient,\cite{Kim:2015el,PhysRevLett.122.057204} though this could
also be attributed to the direct magnon forces in the interacting
Lagrangian.

The first term on the right-hand side of
Eq.~\eqref{eq:lagrangian-cc-simpl} is the skyrmion's kinetic energy
term. The mass tensor of antiferromagnetic textures in the absence of
spin wave excitations has been derived previously by Tveten \emph{et
  al.}~in Ref.~\onlinecite{Tveten:2013aa}. Our solution
agrees\footnote{The unusual $\sqrt 2$ factor in our mass tensor
  compared to that of Tveten \emph{et al.}~arises from our direct use
  of $S=s\hbar$ in our Berry phase Lagrangian; Tveten \emph{et
    al.}~base their notation directly on the gyromagnetic factor.}
with theirs when the magnon number density vanishes, $\rho_n=0$. When
$\rho_n\neq 0$, the form of the mass tensor in
Eq.~\eqref{eq:lagrangian-cc-simpl} indicates that the presence of spin
waves in a skyrmion (or any other spin texture described by rigid
coordinates) will lower the skyrmion's effective mass.

This mass reduction scales with the number and frequency of spin
waves, that is, with the total spin wave energy present in the
skyrmion. We predict that this reduction should occur in any
antiferromagnetic skyrmion system, including those driven by electric
current, as skyrmion motion itself produces spin waves. This mass
reduction suggests skyrmion mobility should be increased in higher
temperature systems, where thermal magnon populations exist throughout
the system. This is consistent with findings in the literature
\cite{Kim:2015el,Barker:2016fq} that the diffusion constant for an
antiferromagnetic skyrmion increases with temperature, though any
precise discussion of the relationship between magnon-driven mass
reduction and diffusive behavior, along with the relative
contributions of the mass reduction and the static potential term on
the far right of Eq.~\eqref{eq:lagrangian-cc-simpl} is left to future
research.

\subsection{Frequency-dependent kinematics}
\begin{figure}
\centering
  \includegraphics[width=\columnwidth]{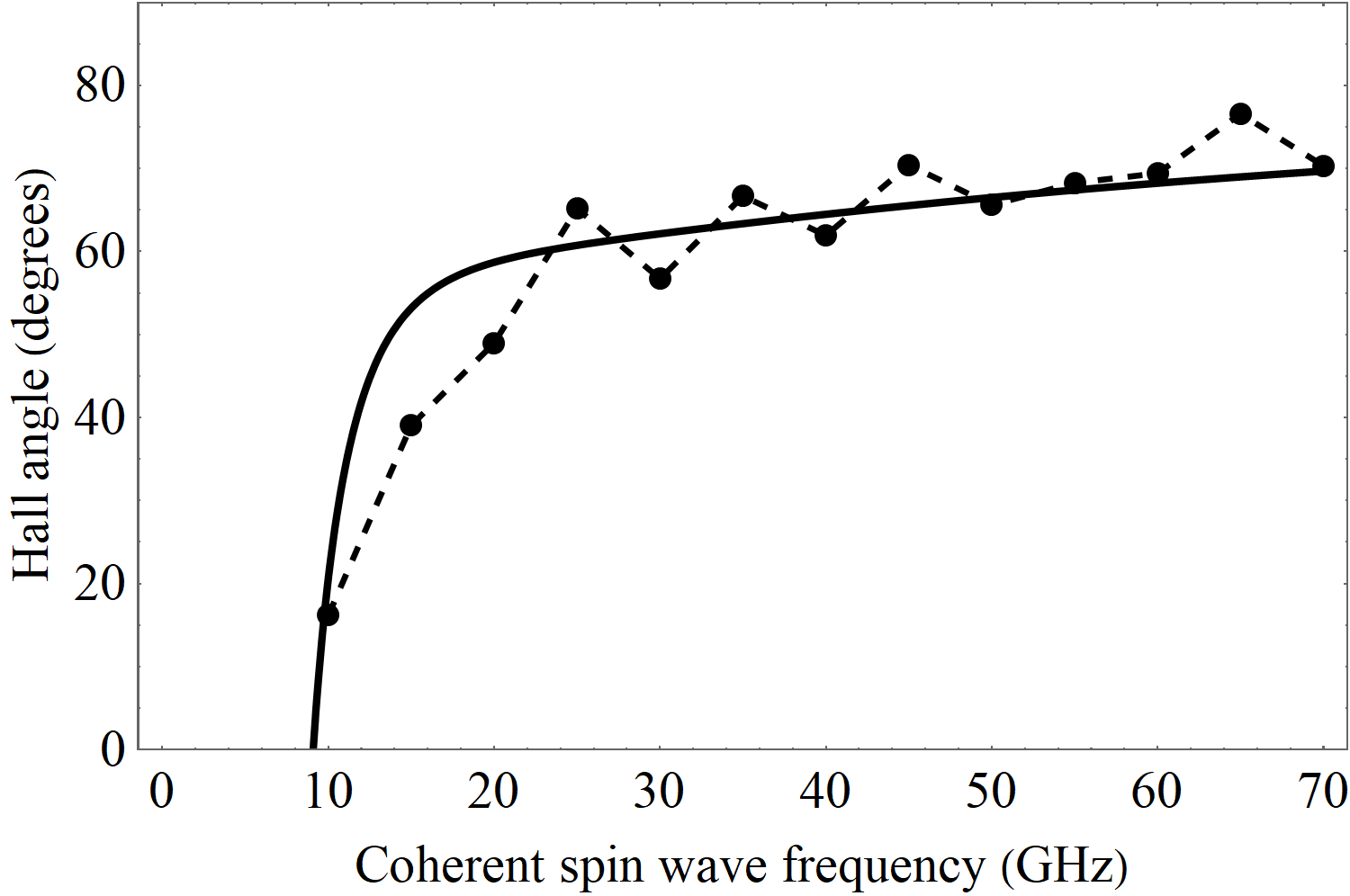}
  \caption{Simulation and theory results for the skyrmion Hall angle
    driven by chirally polarized spin waves. Dashed lines connect the
    data from micromagnetic simulation; the solid line gives the
    theoretical prediction according to Eq.~\eqref{eq:3}. Skyrmion
    position was tracked by a position-weighted expectation value of
    the skyrmion charge density, that is,
    Eq.~\eqref{eq:skyrmion-num-def} taken with a factor of
    $\mathbf{x}$ in the integrand.}
  \label{fig:frequency-dep}
\end{figure}
We have seen that the sign and amplitude of the skyrmion Hall angle is
tunable by changing the chirality of the driving magnon
current. Although ferromagnetic skyrmion systems cannot change the
sign of this angle, the amplitude is known to depend on the frequency
of the driving spin waves.\cite{Schutte:2014aa,Iwasaki:2014aa} In
these systems, the Hall angle increases monotonically with spin wave
frequency until reaching a critical frequency, past which the Hall
angle decays to zero as frequency continues to increase. We have
indicated this ferromagnetic behavior schematically with the dashed
line in Fig.~\ref{fig:frequency-dep}.

We investigated the frequency dependence of the skyrmion Hall angle in
simulation, and present our results in
Fig.~\ref{fig:frequency-dep}. These results (solid line and points)
indicate a strong dissimilarity to the ferromagnetic behavior at high
frequencies. The lack of a critical frequency for optimal Hall
deflection is an attractive property for skyrmionic devices, as one
can achieve uniform angular propulsion without the need to fine tune
the driving frequencies.

The origin of this divergence from the ferromagnetic physics lies,
essentially, in the magnon dispersion relation. As the frequency of a
ferromagnetic spin wave increases, so does its group
velocity. Modeling its interaction with the skyrmion as that of a
charged particle passing through a region of magnetic flux, we can
understand the high frequency decay of the ferromagnetic skyrmion Hall
angle as an explosion of the magnon cyclotron radius at high
velocity.\cite{Iwasaki:2014aa}

In antiferromagnetic systems, the well-known linear
dispersion\cite{kittel:2005aa} at high wavenumber ensures that, away
from the band bottom, group velocity does \emph{not} increase with
increasing frequency. As we move away from the band bottom where
anisotropy induces a locally parabolic dispersion, the group velocity
of the magnon saturates, and so the magnon cyclotron radius and
corresponding skyrmion Hall angle do as well.

\subsection{Flux disk model}

To model the frequency dependence using our theoretical framework, we adopt the simplifying assumption that the skyrmion can be modeled as a disk of uniform flux. The $4\pi Q$ emergent magnetic flux that a real skyrmion would distribute according to the profile of $\bm B$ is instead distributed uniformly over a disk of radius $R$, the skyrmion radius. This approximation has been used in past analyses of magnon-skyrmion scattering with great success.\cite{Iwasaki:2014aa}

Further assuming that the magnitude of the group velocity does not
change over the course of its passage through the skyrmion, the magnon
scattering problem reduces to understanding the intersection of two
circles: the skyrmion circumference and the cyclotron orbit of the
magnon wavepacket, which we define to have radius $r$. We then wish to
analyze the change in direction accumulated by a magnon entering the
skyrmion from the $-\hat y$ direction. The problem enjoys a helpful
constraint: the center of the cyclotron orbit has the same $y$
coordinate as the point of entry $(x,y)$ to the flux disk. It is an
exercise in planar geometry to then show that the angle subtended by
the cyclotron orbit restricted to the flux disk is given by
$\sin(\theta/2) = \sqrt{1-\xi^2}/\sqrt{1+\rho^2+2\rho \xi}$, where
$\rho = r/R$ and $\xi = x/R$. The angle between the
$\Delta \dot{\bm{x}}$ induced by this cyclotron motion and the
$\hat y$ axis is just $\Theta_{SH}=(\pi - \theta)/2$, which will also
be the skyrmion Hall angle by momentum conservation.

The problem of determining a closed expression for the skyrmion Hall
angle therefore reduces to finding the normalized cyclotron radius
$\rho(\xi)$. Choosing the Lorentz force as our lone centripetal force,
we have $\rho = mvR/4\chi Q$ where we have set the magnetic flux
density to $B=4\pi Q/R^2$. Now, supposing that the incoming magnon
current is distributed uniformly along the $\hat x$ direction, the
mean outgoing angle is
\begin{equation}
  \label{eq:3}
  \langle \theta \rangle = \frac{1}{2}\int_{-1}^1\theta(\xi)\,d\xi = \frac{\pi}{2\rho} = \frac{2\pi\chi Q}{m v R},
\end{equation}
which formally converges only in the case $\rho>1$. Finally, from
Eqs.~\eqref{eq:semiclassical-eqs}, extract speed
$v = |\partial\omega/\partial\bm k|$ and effective mass
$m = 1 / |\partial^2\omega/\partial\bm k^2|$. Using the spin wave dispersion outside the skyrmion,
\begin{equation}
  \label{eq:4}
  \omega = \sqrt{\left(Jk^2 + Dk+K\right)\left(Jk^2 + Dk + K+Z\right)},
\end{equation}
we plot the Hall angle prediction given by Eq.~\eqref{eq:3} in Fig.~\ref{fig:frequency-dep}.

The remarkable agreement between this simple model and micromagnetic
simulation suggests that the purely electrodynamical modeling of the
magnon-skyrmion interaction caputres most of the important Hall effect
physics. The magnon wavelength ranges from roughly 200 nm at the
lowest frequency to roughly 30 nm at the highest frequency, which is
on the order of the skyrmion radius.

\begin{figure}
  \centering
  \includegraphics[width=\columnwidth]{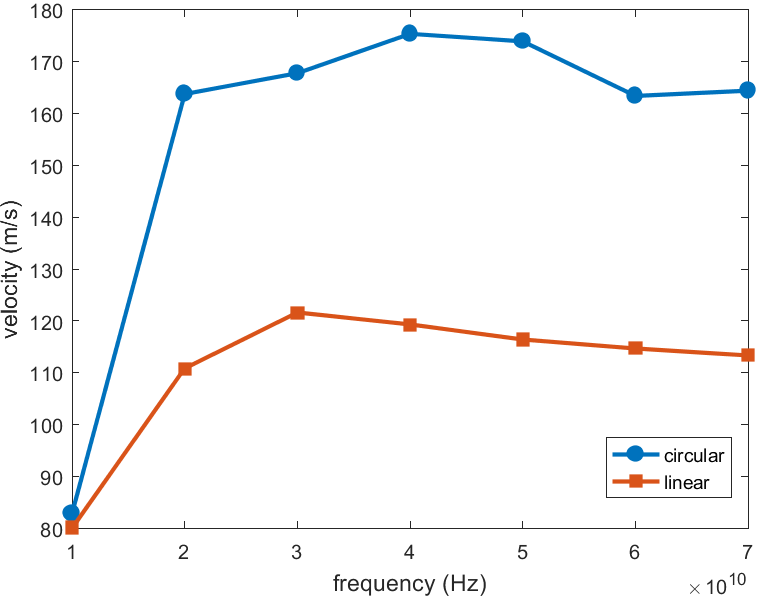}
  \caption{Skyrmion velocity as driven by coherent spin wave signals,
    of either circular or linear polarization, driven by circularly or
    linearly polarized magnetic fields of equal amplitude. The plotted
    frequency range corresponds to the same dynamical range above the
    resonant frequency as in Fig.~\ref{fig:frequency-dep}. Skyrmion
    position was tracked by a position-weighted expectation value of
    the skyrmion charge density, that is,
    Eq.~\eqref{eq:skyrmion-num-def} taken with a factor of
    $\mathbf{x}$ in the integrand.}
  \label{fig:skyrmion-kinematics}
\end{figure}
\begin{figure*}
  \centering
  \includegraphics[width=0.32\textwidth]{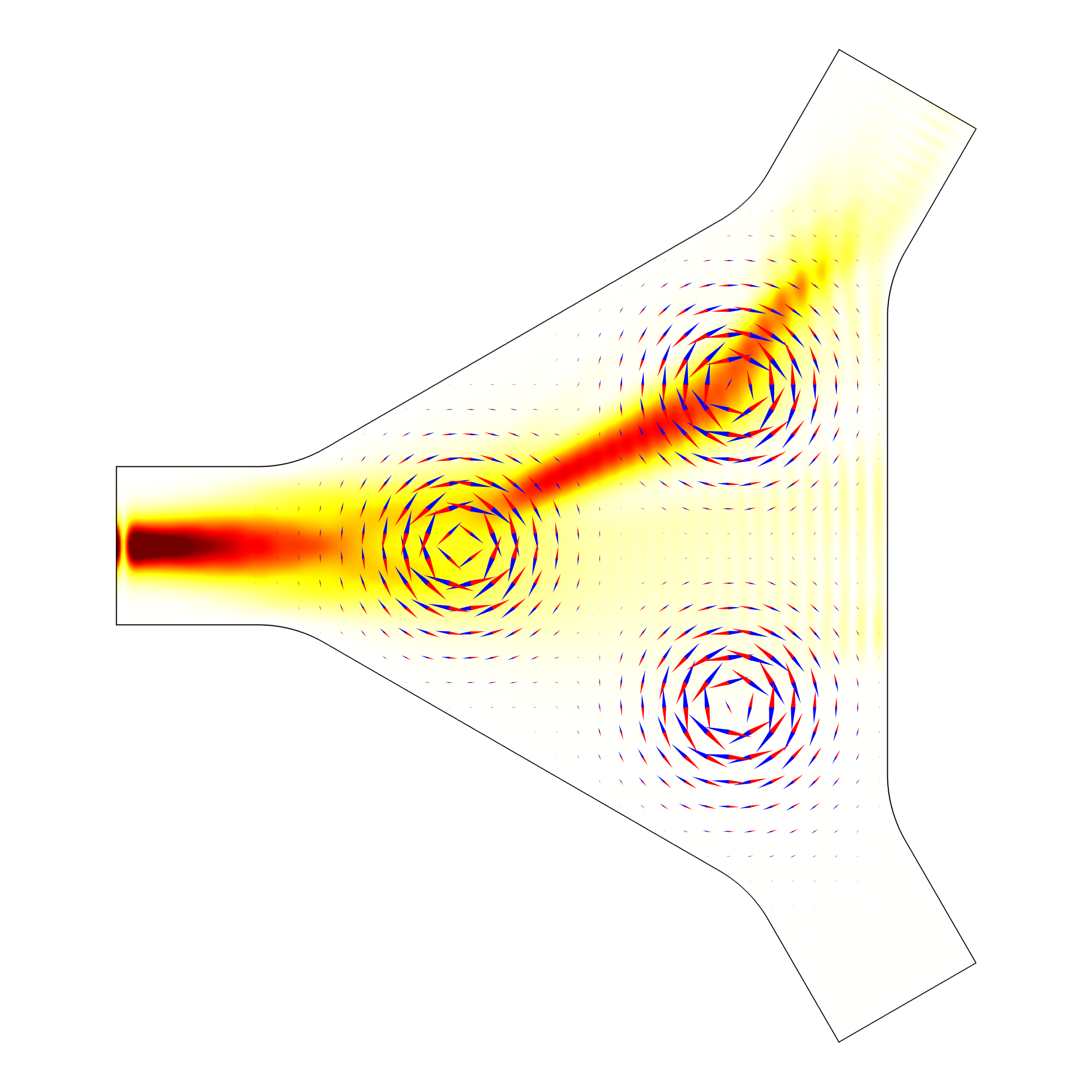}
  \includegraphics[width=0.32\textwidth]{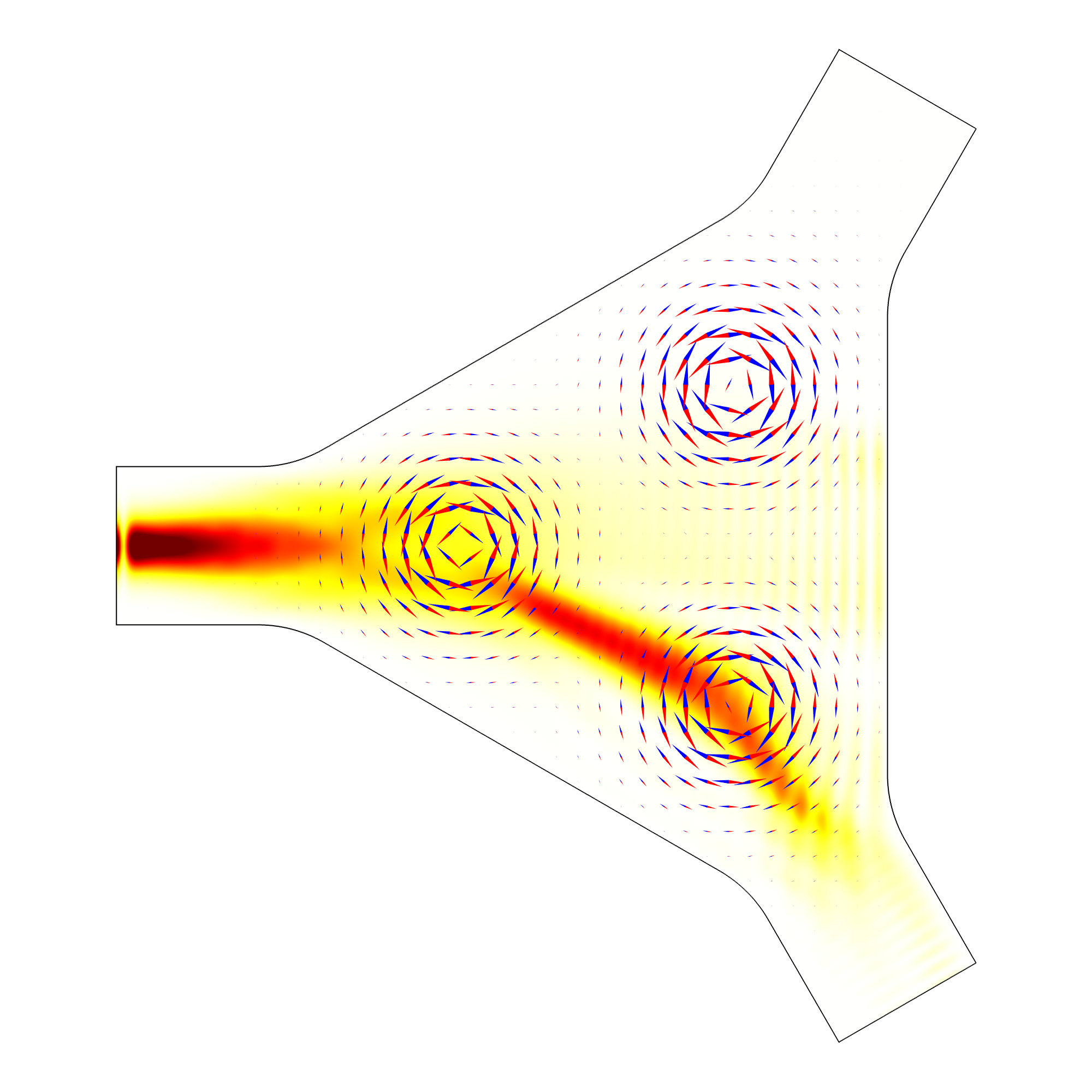}
  \includegraphics[width=0.32\textwidth]{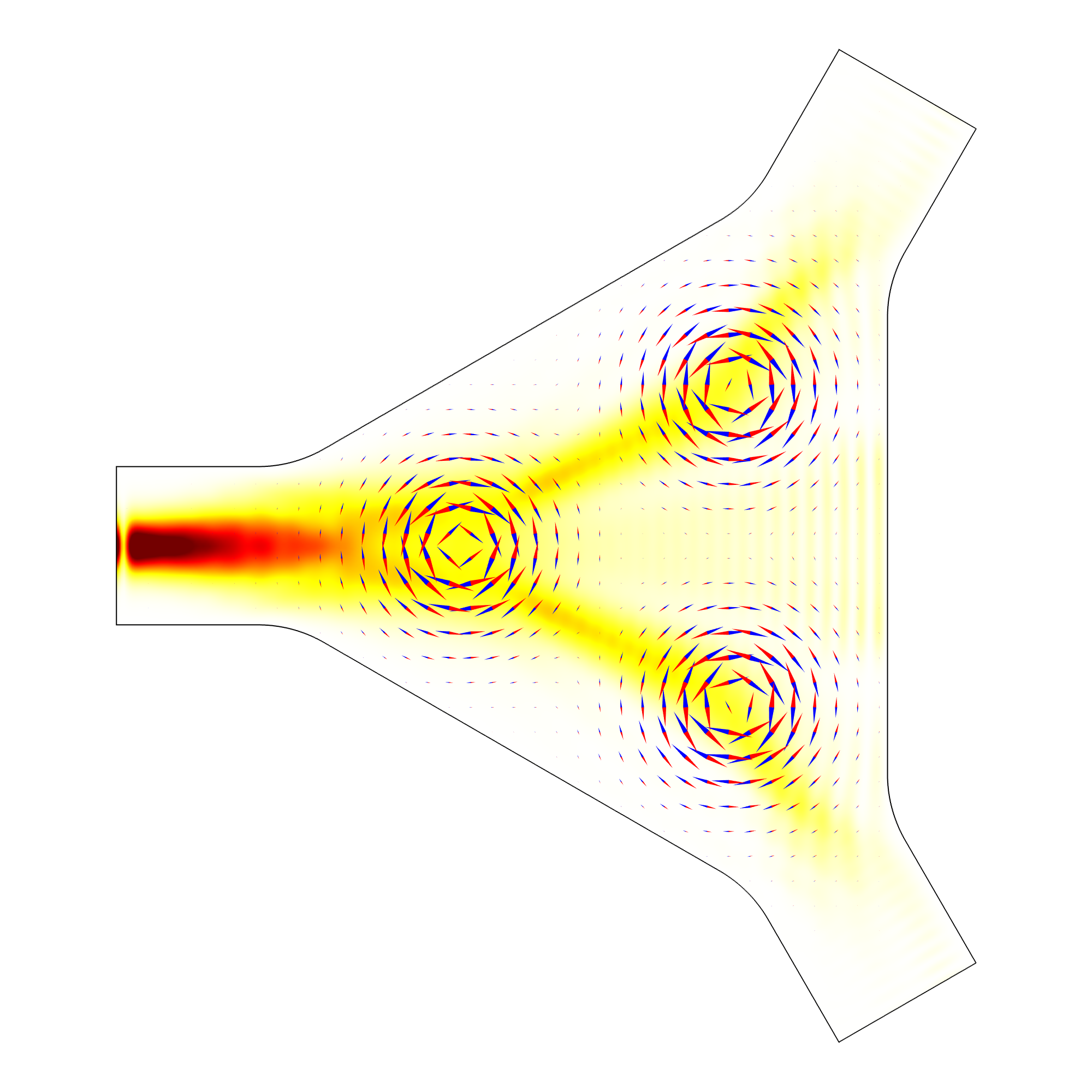}
  \caption{Spin wave circulators implemented using antiferromagnetic
    skyrmions. In all cases, a spin wave current is injected from the
    left. Spin polarized magnon currents circulate about the device
    with the handedness of their internal isospin. For linearly
    polarized waves, the circulator acts as an isolater/filter for the
    spin eigenmodes. The radius from device center to the central edge
    of any port is $200\,\text{nm}$; the injected spin waves have
    frequency $0.2\,\text{THz}$. The color scale represents the square
    of the spin wave amplitude integrated over the full simulation
    time.}
  \label{fig:circulators}
\end{figure*}

In Fig.~\ref{fig:skyrmion-kinematics}, we plot the skyrmion speed as
a function of driving frequency. At reasonable frequencies, the
skyrmion reaches speeds on the order of $100\,\text{m/s}$, comparable
with fast electron-driven ferromagnetic skyrmions. These speeds are also comparable to those predicted for antiferromagnetic skyrmions driven by a thermal gradient.\cite{PhysRevB.99.054423} In the same system
with the same parameters, driving our skyrmion with a current-induced
spin-transfer torque gave speeds on the order of $1000\,\text{m/s}$,
consistent with the literature.\cite{Zhang:2016kf} Though the magnonic drive
cannot reach the speeds of an electronic drive in antiferromagnets, it
readily competes with high speed ferromagnetic systems and could
provide a feasible low-energy route to antiferromagnetic skyrmionics.

That the linearly polarized modes produce a slower speed can be
partially attributed to the competition between right- and left-handed
waves. In the circular spin wave case, the change of momentum for each
spin wave component is unimodal and strongly concentrated around that
mode, as in the left and right panes of
Fig.~\ref{fig:magnon-hall-figs}. The momentum transferred to the
skyrmion will be proportional to the total $\Delta \bm k$ vector lost
by these spin waves, which will be proportional to $2\Theta_{SH}-\pi$.

In the linearly polarized case, depicted in the central pane of
Fig.~\ref{fig:magnon-hall-figs}, horizontal components of the magnonic
force on the skyrmion cancel each other. As we discussed at length in
the beginning of the manuscript, this cancellation provides for the
observed absence of a skyrmion Hall effect in many previous
investigations of antiferromagnetic skyrmion dynamics. This is a
preferable quality of many skyrmion based devices, but we see here
that it comes at a cost. The cancellation of these opposite spin Hall
channels also wastes some of the momentum transfer from magnon to
skyrmion. 

If this Lorentz force were the lone mechanism responsible for this
behavior, then we would expect $v_L = v_C \cos(2\Theta_{SH})$, where
$v_L$ and $v_C$ are the velocities under linear or circular spin wave
drive. That is to say, we expect the tranverse component to be the
same in each case, with the linear drive losing speed only by virtue
of lacking a transverse component. This relation describes the
relationships of Fig.~\ref{fig:skyrmion-kinematics} well at high
frequencies, but does very poorly at low frequencies. This seems to
suggest that at low frequencies, where the spin wave wavelength is
much bigger than the skyrmion radius, a significant failure mode
arises in the simple analysis we have given above.

The primary failure mode of our analysis as applied to the skyrmion
velocity lies in the fact that we have neglected the conservative
force $-\partial\omega/\partial\br$ in
Eq.~\eqref{eq:semiclassical-eqs}. Though at high frequencies this
force will differ between circularly and linearly polarized currents
only perturbatively at $O(D)$, we expect this missing force to be
significantly ($O(1)$) polarization-dependent at low
frequencies. Studies on antiferromagnetic domain walls have shown that
the transmission spectra of alternately polarized antiferromagnetic
spin waves vary significantly below a critical frequency determined by
the Dzyaloshinskii-Moriya interaction and uniaxial anisotropy at the
center of the domain wall.\cite{Lan:2017aa} Understadning the low
frequency regime therefore requires a truly wave-theoretic analysis;
our wavepacket quasiparticles are not senisble constructions when they
are subject to partial transmission and reflection.

As a result of nontrivial transmission and reflection, spin waves at
sufficently lower energy may also become trapped inside the energy
barrier along the skyrmion's circumference, filling bound magnon
states of the skyrmion texture. We are not aware of any thorough
investigations of these bound states in antiferromagnets, but they are
known to cause significant modifications in the physics of
ferromagnetic skyrmions.\cite{Schutte:2014aa} We have alluded to some
of the effects such states may have in antiferromagnetic skyrmions,
such as the mass reduction explored in
Sec.~\ref{sec:magn-medi-reduct}, but a through study of these
phenomena is beyond the scope of the manuscript. We expect future
research in this area could be extremely interesting, and may differ
significantly from the ferromagnetic physics, for two key
reasons. First, unlike in ferromagnets, the spin carried by the bound
magnon states is not locked to the background texture. Second, the
dynamics of the spin carried by these bound modes could be very rich;
a bound mode initially occupied by an externally driven right-handed
magnon can undergo a dynamical evolution of its
isospin,\cite{daniels:2018nonabelian} and the full effects of such an
evolution are yet unknown. In general, the spectrum of bound modes and
corresponding vibrational modes of the skyrmion are complicated
questions, and only recently are full classifications of these degrees
of freedom beginning to be fully explored in general solitonic
systems.\cite{PhysRevD.98.125010}
\section{Applications}
\label{sec:applications}

The previous sections have laid out theory and simulations of the
skyrmion and topological spin Hall effects in antiferromagnets. In the
present section, we discuss how these phenomena might be put to use as
components in a magnonic or skyrmionic logic device.

Ideas around magnonic logic have been around for some time\cite{serga:2010aa} and are
traditionally considered for implementation in ferromagnets. Though
comparatively young, the enterprise of antiferromagnetic magnonics has potential
advantages. It has been pointed out by many authors that antiferromagnetic
insulators lack stray fields and would avoid cross-talk between tightly packed
devices. The terahertz regime in which antiferromagnetic dynamics operates is
widely regarded as attractive for its speed and its uniqueness among solid state
systems, which generally operate in lower frequency bands.

Another limitation of ferromagnetic magnonics is that it can use only
amplitude and phase to encode information, as these are the only
dynamical degrees of freedom available. Antiferromagnets possess an
additional degree of freedom, the isospin, whose interaction with
skyrmions this manuscript has addressed. Whereas ferromagnetic spin
waves are bound by the background magnetization texture to carry a
particular spin current, antiferromagnetic spin waves are
not. Naturally, this increases the amount of information they can
carry. It also changes the nature of the computation, because the
order of operations on an antiferromagnetic spin wave signal affects
the outcome \cite{daniels:2018nonabelian}. Lebrun et al.~have recently
pointed out that the isospin degree of freedom can also allow us to
distinguish coherent magnon signals from thermal magnon currents,
which will be spin unpolarized in
general.\cite{lebrun:2018:tunable,Kim:2015el} In ferromagnets, of
course, thermal magnons carry the same sign of spin as coherent
magnons.

Because skyrmions couple differently to spin up and spin down magnonic
currents, they could be used in principle as part of a magnonic logic
device. We outlined in Ref.~\onlinecite{daniels:2018nonabelian} how
inversion symmetry breaking, domain walls, and other components of the
antiferromagnetic free energy could be used to realize unitary
rotations of the isospin vectors $\bEta$. The use of a
skyrmion-induced topological spin Hall effect, however, would be quite
different. Since the skyrmion projects a signal into its spin
channels, it acts explicitly as a nonlinear element that cannot be
captured as a unitary transformation. As such, it is also
nonreciprocal: sending an output spin wave back into the skyrmion does
not direct it toward the original input. This notion is exemplified by
a device called a spin wave circulator.

The circulator, more generally, is a nonreciprocal device which can
separate input and output channels. It has been widely used to realize
full-duplex communication in nonmagnonic media: the microwave
circulator\cite{hogan:1952} and acoustic
circulator\cite{fleury:2014:sound} are two examples. Spin-wave
circulators are expected to be critical components of magnonic
computers.\cite{vogt:2014realization} Using the topological spin Hall
effects outlined in this article, we present the design of a spin-wave
circulator based on skyrmions in a synthetic antiferromagnet in
Fig.~\ref{fig:circulators}, where three skyrmions are confined in a
three-terminal structure.

Due to the skyrmion-skyrmion and the skyrmion-edge repulsion, the
skyrmions are stable in such a structure as shown in the micromagnetic
simulation results of Fig.~\ref{fig:circulators}. The three ports of
the circulator are structurally equivalent. The spin channel
projection behavior is shown in Fig.~\ref{fig:circulators}. Incoming
magnons are successively scattered by two skyrmions to either the top
or bottom output ports, depending on their spin. Mixed signals are
split into their two spin components.

\section{Conclusion}

In this article, we explored the Hall effects arising from
magnon-skyrmion interactions in easy-axis antiferromagnetic
insulators. The underlying principle for these effects is the breaking
of mirror plane symmetry across quasiparticle trajectories: in the
magnon case, broken by the skyrmion spin texture; and in the skyrmion
case, broken dynamically by the unbalanced scattering profile of a
circularly polarized magnon current. Unlike the ferromagnetic case
(where the sign of the skyrmion Hall angle cannot be tuned),
antiferromagnetic skyrmions offer a large range of dynamic tunability
that could be put to use in logical or neuromorphic magnonic
systems.

We have explored the magnonic case here, where chiral spin waves could
in principle be generated electronically, optically, or
thermally. However, even spin polarized electron currents can be
produced in antiferromagnetic metals,
\cite{PhysRevB.73.214426,PhysRevLett.100.226602} giving unbalanced
spin tranfer torques on the sublattices and breaking the mirror
symmetry we discussed in Sec.~\ref{sec:symmetry-considerations} to
produce Hall physics. Even when the skyrmion does not move, the spin
Hall effect it produces could be useful in such computing schemes, or
as an experimental method for inferring a skyrmion's existence.

We emphasize that although our micromagnetic simulations focused on
synthetic antiferromagnets, our theory is equally applicable to both
synthetic and traditional, bipartite antiferromagnets. Whereas the
former are easily probed in experiment, observing skyrmion-scale spin
textures in antiferromagnets can be a challenge. Using spin Hall
signals as an indirect means of observation may be a useful tool both
experimentally and in skyrmion-based devices. The parameters we used
in our micromagnetic simulations correspond approximately to two
antiferromagnetically coupled layers of yttrium iron garnet (YIG), as
have been used in a variety of similar
studies.\cite{wang:2011:all-magnonic,lan:2015:diode} Recently,
experimental work has demonstrated antiferromagnetic coupling between
ultrathin bilayers of yttrium iron garnet and gadolinium iron
garnet (GdIG).\cite{PhysRevApplied.10.044046} This realization of an
insulating, two-dimensional synthetic antiferromagnet would be an
ideal system for testing the topological spin Hall effect we described
in this paper.

M.W.D.~and W.Y.~contributed equally to this work. We would like to
thank Xiaochuan Wu, Jin Lan, and Mark D.~Stiles for insightful
discussions. This work was supported by the Cooperative Research
Agreement between the University of Maryland and the National
Institute for Standards and Technology, award 70NANB14H209, through
the University of Maryland (M.W.D.), the NSF East Asia and Pacific
Summer Institute under award number EAPSI-1515121 (M.W.D.), the China
Postdoctoral Science Foundation, project number 2018M641906 (W.Y.),
the National Natural Science Foundation of China project number
11847202 (W.Y.), the National Natural Science Foundation of China
grant number 11722430 (J.X.), and the Defense Advanced Research
Project Agency (DARPA) program on Topological Excitations in
Electronics (TEE) under grant number D18AP00011 (D.X.).

%

\end{document}